\documentclass[review,12pt]{elsarticle}

\usepackage{amssymb}
\usepackage{amsthm}
\usepackage{amsmath}
\usepackage{mathrsfs}
\usepackage{graphicx}
\usepackage{epstopdf}
\usepackage{float}
\usepackage{caption}
\usepackage{subcaption}
\usepackage{bm}
\usepackage{bbm}
\usepackage{mathrsfs}
\usepackage{cleveref}
\usepackage{soul}
\usepackage{color,soul} 
\usepackage{color} 
\usepackage{siunitx}
\usepackage[margin=2.5cm]{geometry}
\biboptions{sort&compress}

\journal{Composites Science and Technology}

\makeatletter
\def\@author#1{\g@addto@macro\elsauthors{\normalsize%
    \def\baselinestretch{1}%
    \upshape\authorsep#1\unskip\textsuperscript{%
      \ifx\@fnmark\@empty\else\unskip\sep\@fnmark\let\sep=,\fi
      \ifx\@corref\@empty\else\unskip\sep\@corref\let\sep=,\fi
      }%
    \def\authorsep{\unskip,\space}%
    \global\let\@fnmark\@empty
    \global\let\@corref\@empty  
    \global\let\sep\@empty}%
    \@eadauthor={#1}
}
\makeatother

\begin{document}

\begin{frontmatter}



\title{Phase field predictions of microscopic fracture and R-curve behaviour of fibre-reinforced composites}


\author{Wei Tan\fnref{QMUL}}

\author{Emilio Mart\'{\i}nez-Pa\~neda\corref{cor1}\fnref{IC}}
\ead{e.martinez-paneda@imperial.ac.uk}

\address[QMUL]{School of Engineering and Materials Science, Queen Mary University London, Mile End Road, London, E1 4NS, UK}

\address[IC]{Department of Civil and Environmental Engineering, Imperial College London, London SW7 2AZ, UK}

\cortext[cor1]{Corresponding author.}

\begin{abstract}
We present a computational framework to explore the effect of microstructure and constituent properties upon the fracture toughness of fibre-reinforced polymer composites. To capture microscopic matrix cracking and fibre-matrix debonding, the framework couples the phase field fracture method and a cohesive zone model in the context of the finite element method. Virtual single-notched three point bending tests are conducted. The actual microstructure of the composite is simulated by an embedded cell in the fracture process zone, while the remaining area is homogenised to be an anisotropic elastic solid. A detailed comparison of the predicted results with experimental observations reveals that it is possible to accurately capture the crack path, interface debonding and load versus displacement response. The sensitivity of the crack growth resistance curve (R-curve) to the matrix fracture toughness and the fibre-matrix interface properties is determined. The influence of porosity upon the R-curve of fibre-reinforced composites is also explored, revealing a stabler response with increasing void volume fraction. These results shed light into microscopic fracture mechanisms and set the basis for efficient design of high fracture toughness composites. \end{abstract}


\end{frontmatter}



\section{Introduction}
\label{Introduction}

Lightweight fibre reinforced polymer (FRPs) composites are being widely used in aeronautical and automotive applications due to their high specific stiffness and strength. To meet the structural integrity requirements of composite structures used in transportation vehicles, FRPs need to be sufficiently damage-tolerant to sustain defects safely until they can be repaired \cite{Tan2015}. This requires composite structures of high fracture toughness, a property that depends on the mechanical properties of fibre, matrix, and fibre-matrix interfaces, as well as of their spatial distribution within the material.

The fracture of FRPs can be generally classified into two categories: interlaminar fracture (delamination) and intralaminar fracture. Interlaminar fracture toughness values are controlled by the matrix toughness, which typically ranges from 0.1 to 3 kJ/m\textsuperscript{2} \cite{Cowley1997,Tan2016}. Intralaminar fracture can be classified into two categories, namely fibre-dominated fracture and matrix-dominated fracture. The reported intralaminar fracture toughness FRPs are in the range of 1 to 634 kJ/m\textsuperscript{2} \cite{Tan2016,Laffan2012,Marin2016}. The matrix-dominated fracture toughness is comparable to interlaminar fracture toughness ($\sim$1 kJ/m\textsuperscript{2}), while the fibre-dominated fracture toughness is two order of magnitude higher. This is primarily due to the fibre-bridging effect, where a significant amount of fracture energy is absorbed by fibre-matrix debonding, fibre pull-out and fibre breakage.  

Avenues for improving composite fracture toughness include matrix modification, thermoplastic particles, nanomaterial veils, stitching, Z-pin and 3D fibre architectures. These methods generally take advantage of well-known toughening mechanisms such as crack deflection, microcrack toughening, fibre/grain bridging.  While trial-and-error experimental techniques are available to improve the fracture toughness, another emerging approach is the application of computational micromechanics. This approach is based on the finite element simulation of the mechanical response of a representative volume element (RVE) or an embedded cell of the composite microstructure. This makes possible to (virtually) optimise the material properties by changing the properties of the constituents. It can also provide the homogenized constitutive behaviour of the composite material, which can then be transferred to simulations at a larger length scale \cite{Llorca2011,Tan2018,Herraez2018}.

Cohesive Zone Models (CZM) \cite{Camanho2002,Canal2012} and Continuum Damage Mechanics (CDM) models  \cite{Chaboche1988a, Tan2015} are being extensively used in computational micromechanics. However, one source of mesh-dependence in CDM or CZM models is the mesh-induced direction bias. The misalignment between crack band direction and mesh lines induces stress locking because of the displacement continuity condition. A practical solution to mitigate mesh-induced directional bias is to align a refined mesh with the fibre direction \cite{Falco2018}, requiring complex mesh generations and a high computational cost. To overcome this issue, different element enriched formulations have been proposed, such as the eXtended FEM (X-FEM) \cite{Belytschko2009} and the Floating Point Method \cite{Chen2014a}. Despite their effectiveness, these techniques can also fail to track the actual crack path topology, whereby crack coalescence and branching scenarios may potentially occur. A promising alternative for modeling the progressive failure of materials is the Phase Field (PF) fracture model \cite{Bourdin2000,Miehe2010a,TAFM2020}, which is gaining a growing interest in the scientific community \cite{Wu2020}. In particular, this approach enables to accurately simulate complex crack paths, including crack branching and coalescence in arbitrary geometries and dimensions. The PF method is a variational approach to fracture that exploits the classical Griffith energy balance \cite{Griffith1920}; cracking takes place when the energy released by the solid reaches a critical value, the material toughness $G_c$. Recently, Quintanas-Corominas et al. \cite{Quintanas-Corominas2018,Quintanas-Corominas2019,Quintanas-Corominas2019a, Quintanas-Corominas2020a} and Espadas-Escalante et al. \cite{Espadas-Escalante2019} have successfully used the PF model to capture the intralaminar and interlaminar damage behaviours at the mesoscale level. However, important phenomena governing the crack path topology and macroscopic fracture toughness remain unaddressed; these include the influence of fibre, matrix, and fibre-matrix interface, as well as other toughening or embrittlement mechanisms (i.e. fibre bridging, crack branching, voids, defects, etc).  

In this work, a coupled PF-CZM framework is presented to model the matrix cracking, fibre-matrix interface debonding, and homogenised fracture toughness. Finite element modelling of single edge notched three-point bending tests are conducted. The predicted results are validated against the measured crack path and load-displacement curves. The main novel aspects herein are: (i) For the first time, a combined PF-CZM model is used to predict the miscroscale crack propagation and investigate the debonding and matrix bridging behaviour. (ii) The effect of matrix toughness, interface strength and toughness on the crack trajectory and the R-curve are firstly quantified. (iii) We explore the influence on the fracture toughnesss of microstructures with varying degrees of porosity. Our model opens new opportunities for the efficient and cost-effective design of energy-absorbing materials and structures.

\section{Numerical model}
\label{Sec:NumModel}

The formulation combines two fracture models. The phase field fracture method, capable of capturing arbitrary crack trajectories, is used to model crack initiation and growth along the matrix and the fibres. Furthermore, fibre-matrix debonding is simulated using a cohesive zone model. Both models are described below and implemented in the commercial finite element package ABAQUS by means of user element subroutines.

\subsection{Phase field fracture model}

\begin{figure}[h]
    \centering
    \includegraphics[width=1.0\textwidth]{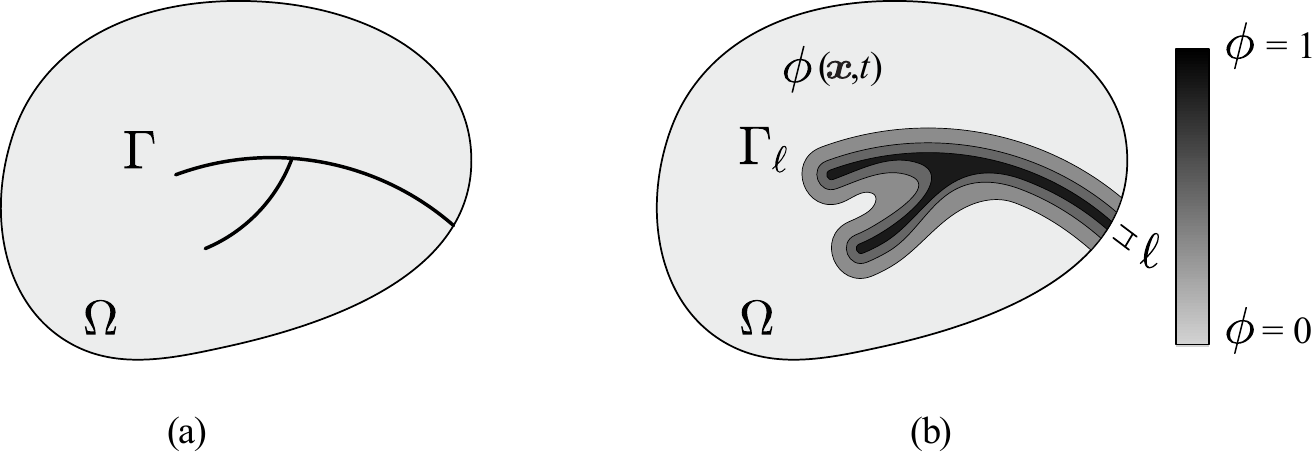}
    \caption{Schematic representation of a solid body with (a) internal discontinuity boundaries, and (b) a phase field approximation of the discrete discontinuities.}
    \label{Fig:PFM}
\end{figure}

The phase field fracture method builds upon Griffith's thermodynamics \cite{Griffith1920}; crack advance is driven by the competition between the work required to create a new surface and the strain energy released by the solid as the crack grows. Griffith's energy-based failure criterion can be expressed in variational form \cite{Francfort1998}. Thus, consider an arbitrary body $\Omega \subset {\rm I\!R}^n$ $(n \in[1,2,3])$ with internal discontinuity boundary $\Gamma$. The total potential energy of the body will be a sum of the contributions associated with the strain energy density $\psi$ and the fracture energy $G_c$ as,
\begin{equation}\label{eq:Egriffith}
    \mathcal{E} \left( \bm{u} \right) = \int_\Omega  \psi \left( \bm{\varepsilon} \left( \bm{u} \right) \right) \, \text{d}V +  \int_\Gamma  G_c \, \text{d}S \, ,
\end{equation}

\noindent where $\bm{u}$ and $\bm{\varepsilon}=\left( \nabla \bm{u}^T + \nabla \bm{u} \right)/2$ denote the displacement and strain fields, respectively. Minimisation of the Griffith energy functional (\ref{eq:Egriffith}) is hindered by the complexities associated with tracking the propagating fracture surface $\Gamma$. However, an auxiliary variable, the phase field $\phi$, can be used to track the crack interface; $\phi$ is a damage-like variable that goes from 0 in intact regions to 1 inside of the crack - see Fig. \ref{Fig:PFM}.

Following continuum damage mechanics arguments, a degradation function $g=(1-\phi)^2$ is defined that diminishes the stiffness of the material with evolving damage. Accordingly, the total potential energy functional can be re-formulated as
\begin{equation}
    \mathcal{E}_\ell \left( \bm{u}, \phi \right) = \int_\Omega \left( 1 - \phi \right)^2 \psi \left( \bm{\varepsilon} \left( \bm{u} \right) \right) \, \text{d}V + \int_\Omega G_c  \left( \frac{\phi^2}{2 \ell} + \frac{\ell}{2} |\nabla \phi|^2 \right) \, \text{d}V \, ,
\end{equation}

\noindent where $\ell$ is a length scale parameter that governs the size of the fracture process zone; the non-local character of the phase field method guarantees mesh objectivity. As rigorously proven using Gamma-convergence, the $(\bm{u}, \phi)$ sequence that constitutes a global minimum for the regularised functional $\mathcal{E}_\ell$ converges to that of $\mathcal{E}$ for a fixed $\ell \to 0^+$. Thus, $\ell$ can be interpreted as a regularising parameter in its vanishing limit. However, for $\ell>0^+$ a finite material strength is introduced and $\ell$ becomes a material property governing the strength \cite{Tanne2018}; e.g., for plane stress:
\begin{equation}
  \sigma_f  \propto \sqrt{\frac{G_c E}{\ell}} = \frac{K_{Ic}}{\sqrt{\ell}}
\end{equation}
\noindent where $K_{Ic}$ is the material fracture toughness.

Finally, the strong form can be readily derived by taking the first variation of $\mathcal{E}_\ell$ with respect to the primal kinematic variables and making use of Gauss' divergence theorem. Thus, the coupled field equations read, 
\begin{align}\label{eqn:strongForm}
(1-\phi)^2 \, \, \nabla \cdot \boldsymbol{\sigma}  &= \boldsymbol{0}   \hspace{3mm} \rm{in}  \hspace{3mm} \Omega \nonumber \\ 
G_{c}  \left( \dfrac{\phi}{\ell}  - \ell \Delta \phi \right) - 2(1-\phi) \, \psi  &= 0 \hspace{3mm} \rm{in} \hspace{3mm} \Omega  
\end{align}

\noindent The discretised forms of the field equations are solved by using a staggered solution scheme \cite{Miehe2010a,CPB2019}. 

\subsection{Cohesive zone model} 

\begin{figure}[h]
    \centering
    \includegraphics[width=1.0\textwidth]{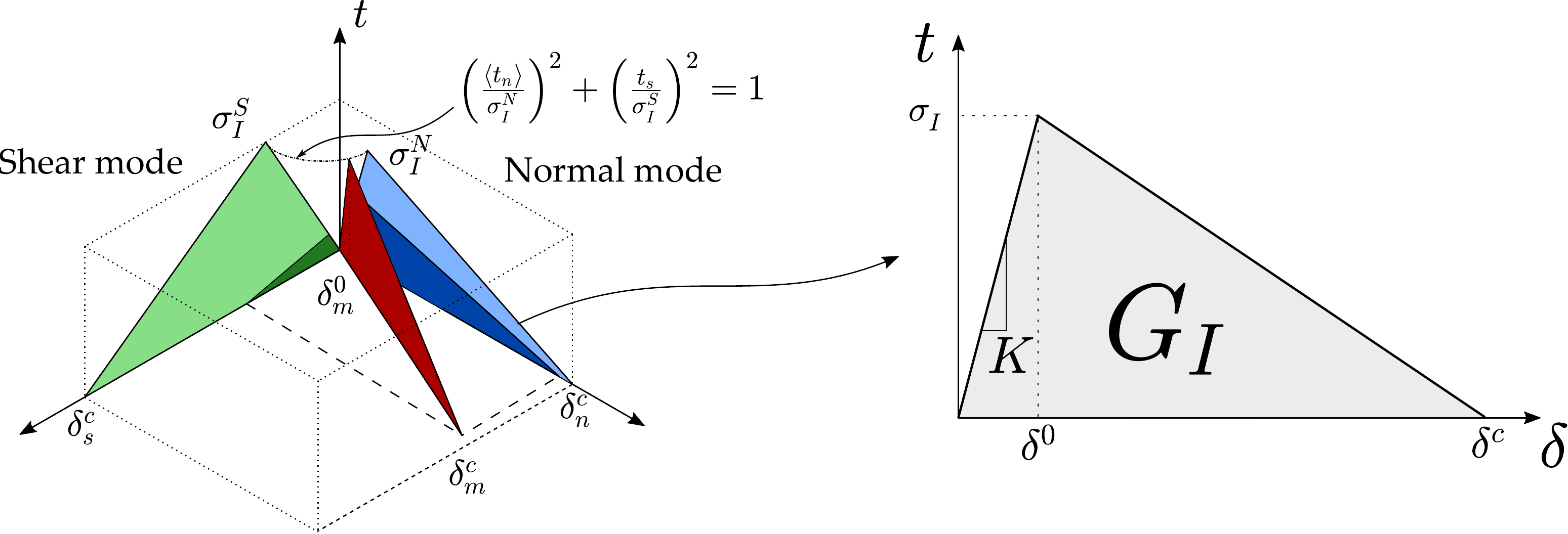}
    \caption{Sketch of the cohesive zone formulation employed for predicting fibre-matrix debonding.}
    \label{Fig:CZM}
\end{figure}

Debonding between the matrix and the fibre is captured by means of a cohesive zone model with a bi-linear traction-separation law, as shown in Fig. \ref{Fig:CZM}. For both normal and shear tractions, the constitutive behaviour of the cohesive zone interface is governed by the initial interface modulus $K$, the interface strength $\sigma_I$ and the fracture energy $G_I$.

Following Camanho and Davila \cite{Camanho2002}, an effective separation is introduced to describe the evolution of damage under a combination of normal and shear deformation
\begin{equation}
    \delta_m = \sqrt{\langle \delta_n \rangle^2 + \delta_s^2 }
\end{equation}

The onset of damage is predicted in terms of the normal $t_n$ and shear $t_s$ tractions using a quadratic nominal stress criterion,
\begin{equation}
    \left( \frac{\langle t_n \rangle}{\sigma_I^N} \right)^2 +\left( \frac{ t_s}{\sigma_I^S} \right)^2 =1
\end{equation}

\noindent Finally, damage evolution is governed by the energetic Benzeggagh-Kenane fracture criterion. Thus, the mixed-mode critical energy release rate $G_C$ will be attained when,
\begin{equation}
    G_I^N+ \left( G_I^S- G_I^N \right) \left( \frac{G^S}{G^N + G^S} \right)^\eta = G_C
\end{equation}

\noindent where $\eta$ is a material parameter, and $G_I^N$ and $G_I^S$ respectively denote the fracture energies required to cause failure in the normal and shear directions.

\section{Results}
\label{Sec:FEMresults}

\subsection{Singe-edge cracked plate subjected to tension}
\begin{figure}[h]
    \centering
    \includegraphics[width=1.0\textwidth]{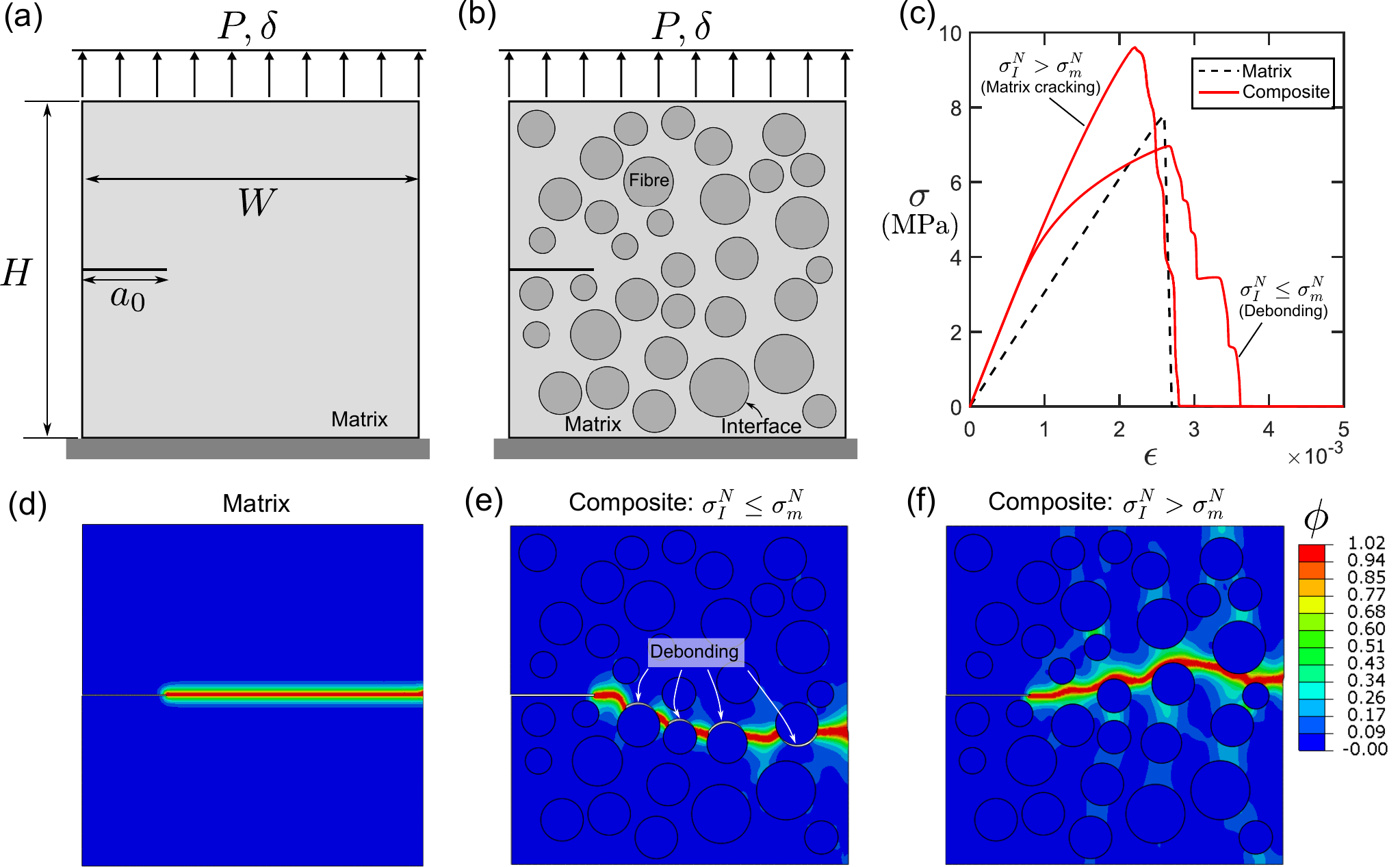}
    \caption{Single-edge cracked plate subjected to tension: (a) Model setup for matrix, and (b) composite. (c) Stress-strain response of single-edge cracked plates made from bulk matrix or fibre-reinforced composite. (d) Crack propagation of matrix. Crack propagation of composite where (e) $\sigma_m^N \leq \sigma_I^N$ and (f) $\sigma_m^N > \sigma_I^N$.}
    \label{Fig.1}
\end{figure}

To verify our PF model on bulk matrix, we model a singe-edge cracked plate with the geometric setup, dimensions and boundary conditions given in Fig. \ref{Fig.1}a. The square plate of width $H=$ 1 mm and height $W=$ 1 mm has an initial crack length of $a_0=$ 0.25 mm. We load the plate by prescribing the vertical displacement in the upper edge, and fix both vertical and horizontal displacements in the bottom boundary. We adopt the following epoxy material properties for the cracked plate, Young's modulus $E=$ 3.5 GPa, Poisson's ratio $\nu=$ 0.35, tensile strength $\sigma_{N}=$ 20 MPa and critical energy release rate $G_m=$ 10 J/m\textsuperscript{2}. 

To assess the effect of fibre reinforcement on the crack propagation of composite material, we use the same geometry, dimensions and boundary conditions as above, except for the additional $f_g=$ 37.2 \% fibre reinforcements, see Fig. \ref{Fig.1}b. We use E-glass fibre of Young's modulus $E=$74 GPa, Poisson's ratio $\nu=$ 0.35 and critical energy release rate of $G_f=$ 13.5 J/m\textsuperscript{2}. Both glass fibre and epoxy matrix are assumed to be linear elastic, isotropic solids. A cohesive surface contact between fibre and matrix is defined and follows a traction-separation law with the properties given in Table \ref{table.1}, where the interfacial tensile strength is assumed to be two-thirds of the shear strength,  $\sigma_I^N=2 \sigma_I^S/3$ \cite{Herraez2018}. To compare the effect of matrix cracking and interface debonding more directly, we choose two sets of material parameters: namely $\sigma_I^N \leq \sigma_m^N  $ and $\sigma_I^N > \sigma_m^N  $.

\begin{table}[H]
    \caption{ Properties of fibre-matrix interface \cite{Herraez2018}} 
    \centering 
    \begin{tabular}{c c c c c c c} 
    \hline\hline 
    $\sigma_I^N$ (MPa) & $\sigma_I^S$ (MPa) & $K^N$ (GPa) & $K^S$  (GPa) & $G_I^N $ (J/m\textsuperscript{2}) & $G_I^S $ (J/m\textsuperscript{2}) & $\eta$\\ [0.5ex] 
    \hline 
    40 & 60 & 1000 & 1000 & 125 & 150 & 1.2\\ [1ex] 
    \hline 
    \end{tabular}
    \label{table.1} 
\end{table}

Four-node quadrilateral plane strain elements were used. After a mesh sensitivity study, a fine mesh with a characteristic element size $h=$0.001 mm is used, eight times smaller than the phase field length scale \cite{CMAME2018}. When conducting the mesh sensitivity analysis, attention is paid to ensure that the fracture process zones associated with both the phase field and the cohesive zone model are resolved. In total, 11,221 and 42,361 elements are used for the matrix and composite models, respectively.  

The predicted stress-strain responses of single edge cracked plate made from bulk matrix and fibre-reinforced composites are summarised in Fig. \ref{Fig.1}c. The stress-strain curve of matrix model shows a linear elastic behaviour until reaching the peak load. This is followed by a load drop, corresponding to the crack evolution, see Fig. \ref{Fig.1}d. If fibres are added to the matrix, a stiffening and toughening effect is observed on the overall material behaviour. If the fibre-matrix interface debonding initiates first ($\sigma_I^N \leq \sigma_m^N$), there is a notable non-linear behaviour prior to load drop. Before reaching the peak load, a large number of fibre-matrix interfaces have experienced decohesion, which contributes to the unusual non-linear response. The multi-step load dropping in the softening regime is attributed to the coalescence of interface debonding and matrix cracking, see Fig. \ref{Fig.1}e. However, if the matrix cracking initiates first ($\sigma_I^N > \sigma_m^N$), no interfacial decohesion is observed. A linear elastic behaviour is predicted before the maximum load, followed by a zig-zag softening behaviour. This is mainly due to the crack deflection effect in the fibre-reinforced composites. Instead of a straight cracking trajectory, the crack propagating through the matrix will deflect upon encountering the fibres (Fig. \ref{Fig.1}f), hence increasing the fracture surface area and the macroscopic fracture toughness. To quantify the role of the fibres, we estimate an equivalent work of fracture as the area under the resulting stress-strain curve divided by the ligament crack surface area 0.75 mm\textsuperscript{2}. We find that the composite with interface debonding has the highest work of fracture, 21.7 J/m\textsuperscript{2}, followed by the composite without interface debonding (20.4 J/m\textsuperscript{2}), with the bulk matrix giving 14.3 J/m\textsuperscript{2}. Therefore, to improve the macroscopic fracture toughness of fibre-reinforced composites, the fibre-matrix interface strength should be reduced, consistent with most toughening approaches used in ceramic fibre-reinforced composites \cite{Jiang2018}. However, to improve the strength of fibre-reinforced composites, a high fibre-matrix interface strength is required.  

\subsection{Single-edge notched three-point bending test} 
\begin{figure}[H]
    \centering
    \includegraphics[width=0.9\textwidth]{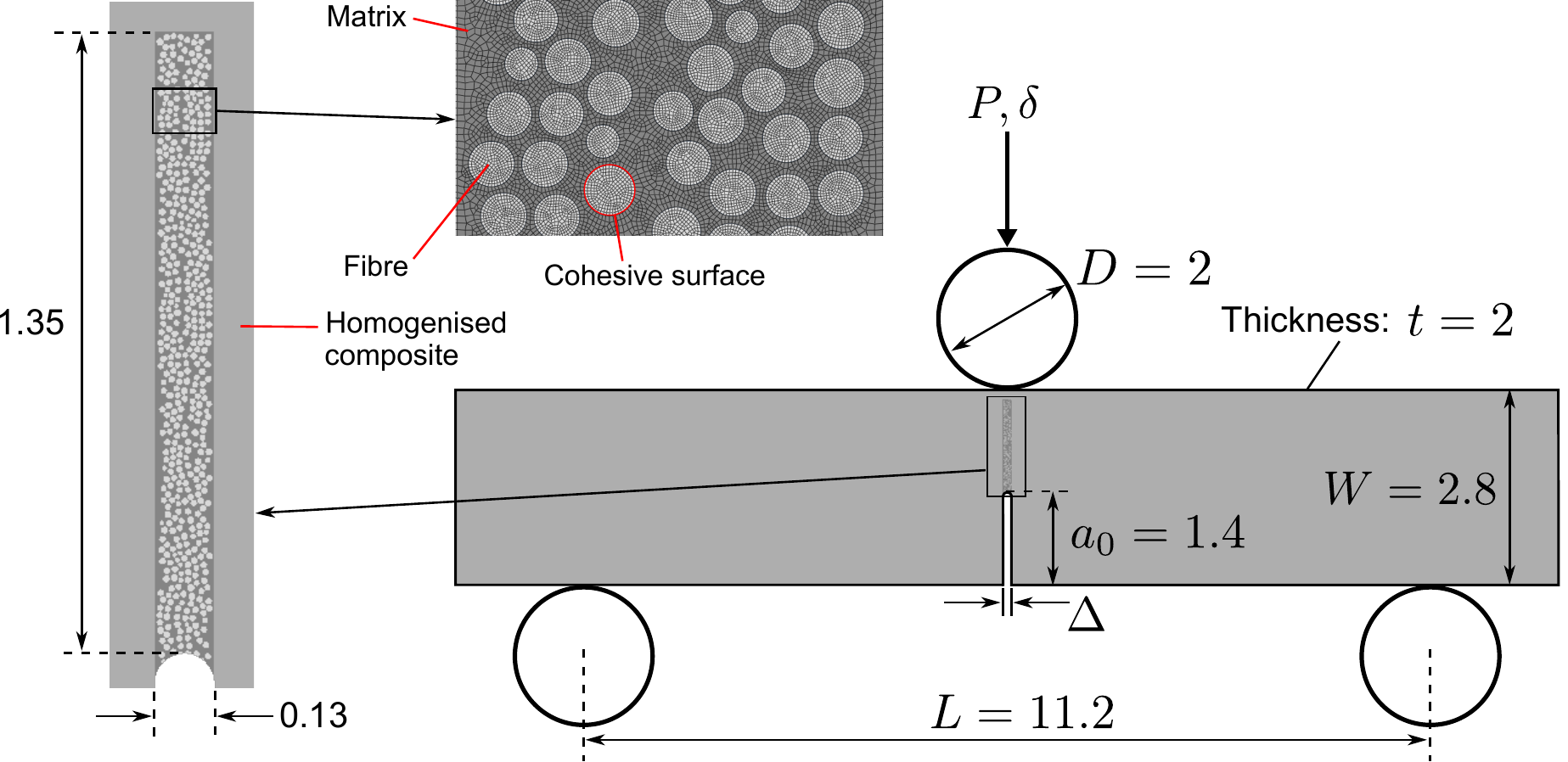}
    \caption{Model set-up of single-edge notched tension tests. All dimensions are in mm.}
    \label{Fig.2}
\end{figure}

We proceed to simulate three-point bending (TPB) experiments on a notched beam to predict the microscale crack topology and the matrix-dominated toughness of the composite lamina. This is achieved by means of an embedded cell model, following the approach developed in \cite{Herraez2018,Canal2012}. As shown in Fig. \ref{Fig.2}, the complete composite microstructure is resolved in the fracture process zone as an embedded cell, while the remaining ply material is represented as a homogeneous, transversely-isotropic elastic solid. The two regions share nodes at their interface, implying a continuous displacement field between the homogenised region and the embedded cell. We calculated the material constants of the homogenised region based on Mori–Tanaka method \cite{Canal2012}. The Young's modulus is $E_h=$ 11 GPa and the Possion's ratio is $\nu_h=$ 0.3. The sample dimensions and experimental setup are given in Fig. \ref{Fig.2}. A single edge-notched beam with a support span $L=11.2$ mm, equal to four times the width $W$, is loaded in three-point bending. The thickness of the beam is $t=$ 2 mm. The initial crack length is $a_0=$ 1.4 mm. Inside the embedded cell, the randomly distributed glass fibres of volume fraction $f_g=$ 54 \% are surrounded by epoxy matrix. Fibre diameter ranges from 13 $\si{\micro\metre}$ to 17 $\si{\micro\metre}$. The characteristic element size is set to 1  $\si{\micro\metre}$ in the embedded region and gradually grows to 0.2 mm at the outer edges. The whole model is formed by 152,364 four-node plane strain elements. The fibre, matrix and fibre-matrix interface properties used in the previous section were taken as baseline input parameters. The applied load P, the loading point displacement $\delta$ and the crack mouth opening displacement (CMOD), $\Delta$, were continuously recorded during the virtual tests. 

\begin{figure}[H]
    \centering
    \includegraphics[width=1.0\textwidth]{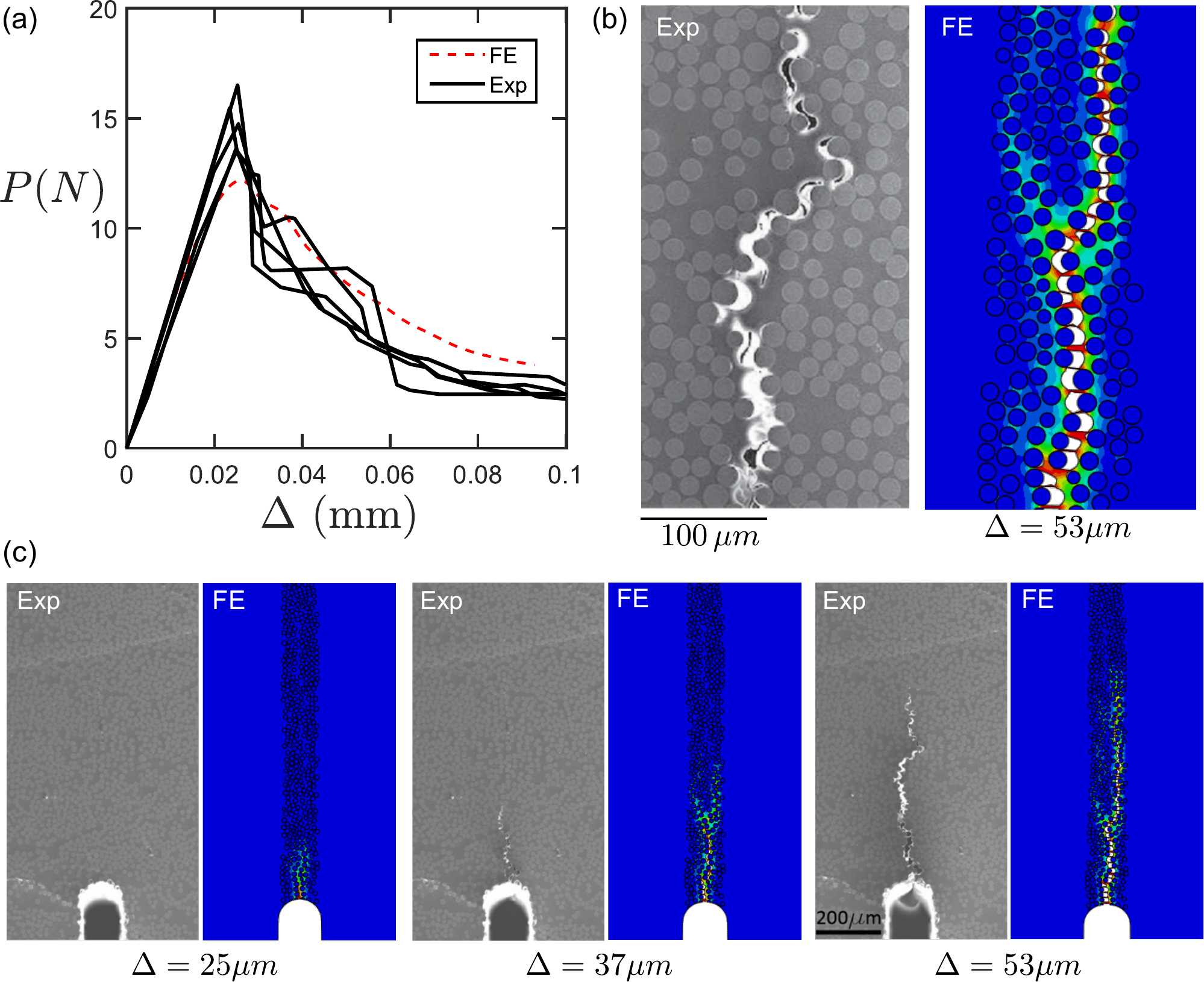}
    \caption{Measured \cite{Canal2012} and predicted (a) load-CMOD curve, (b)crack propagation at high magnification and (c) crack propagation at low magnification.}
    \label{Fig.3}
\end{figure}

The predictions of the virtual three-point bending test are shown in Fig. \ref{Fig.3}. First, the measured \cite{Canal2012} and simulated load-CMOD displacement are plotted in Fig. \ref{Fig.3}a using the embedded cell model presented above. The numerical model accurately captures the measured behavior including the linear-elastic response of the beam before the peak load, the CMOD at the maximum load and the softening regime of the curve. The maximum load is slightly underestimated (around 10\%), within the experimental scatter. In addition to the load-CMOD response, the model is able to reproduce the microscopic deformation and failure mechanisms, see Fig. \ref{Fig.3}b. In agreement with what is observed in the scanning electron micrographs, damage began by interface debonding at the outer surface the fibres. Cracks propagated along the fibre–matrix interface and voids grew by distinct interface separation. A continuous crack path was finally developed by the coalescence of matrix cracking and interface decohesion, while a significant amount of matrix ligaments were bridging the crack.
The numerical simulations also precisely capture the crack evolution with increasing remote load. This is shown in Fig. \ref{Fig.3}c, where snapshots of scanning electron micrographs for different values of the CMOD are plotted together with the predicted results. 

\begin{figure}[H]
    \centering
    \includegraphics[width=1.0\textwidth]{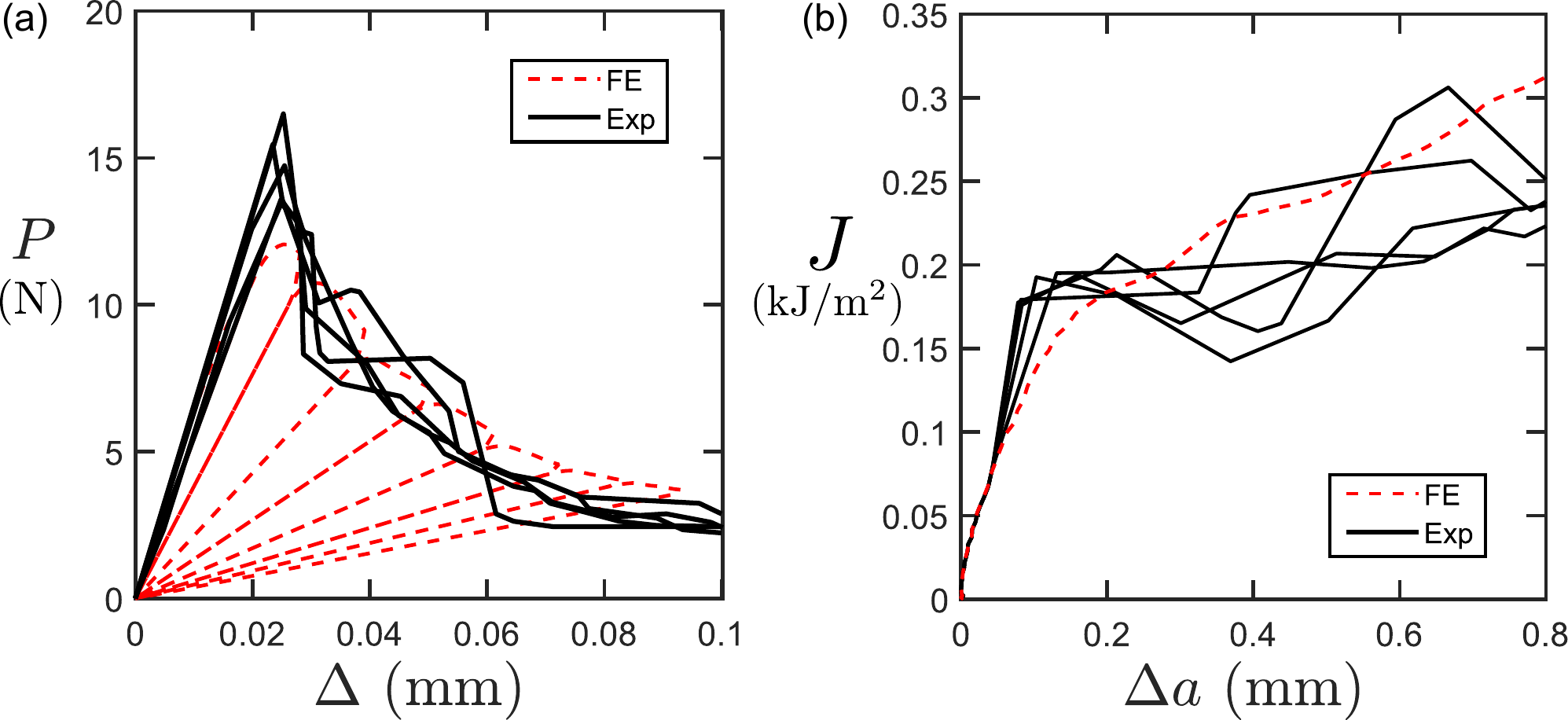}
    \caption{(a) Measured load-CMOD response \cite{Canal2012} is compared with the predicted loading-unloading response (b) Measured \cite{Canal2012} and predicted R-curves.}
    \label{Fig.4}
\end{figure}

Cyclic loading was also applied to the TPB specimen to investigate unloading and reloading behaviour and calculate the unloading compliance $C=\delta/P$. This is facilitated by the linear elastic fracture response of composite materials, as confirmed by the unloading response to the origin shown in Fig. \ref{Fig.4}a - no plastic effects have been considered. Thus, we follow the ASTM standard \cite{ASTM1820} to calculate the R-curve. In brief, the elastic compliance is used to calculate the effective crack size $a_e$, which is then used to calculate the geometrical correction factor $f(a_e/W)$. The stress intensity factor was then given by $K=P S(BW^{3/2})^{-1}f(a_e/W)$. Finally, the $J$-integral is estimated by substituting $K$ into the plain strain equation below,
\begin{equation}\label{eqn.1}
	J=\frac{K^2(1-\nu^2)}{E} \, .
\end{equation}

\noindent The change in $J$ with crack extension determines the R-curve.

The measured and predicted R-curves are plotted in Fig. \ref{Fig.4}b. Predictions for the R-curve response of the TPB test agree closely with those measured in the experiments. The rising R-curves observed both in the experiments and in the numerical predictions are attributed to the bridging effect from the matrix ligaments and the softening behaviour of fibre-matrix interface decohesion. 

\subsubsection{Sensitivity study} 

\begin{figure}[H]
    \centering
    \includegraphics[width=1.0\textwidth]{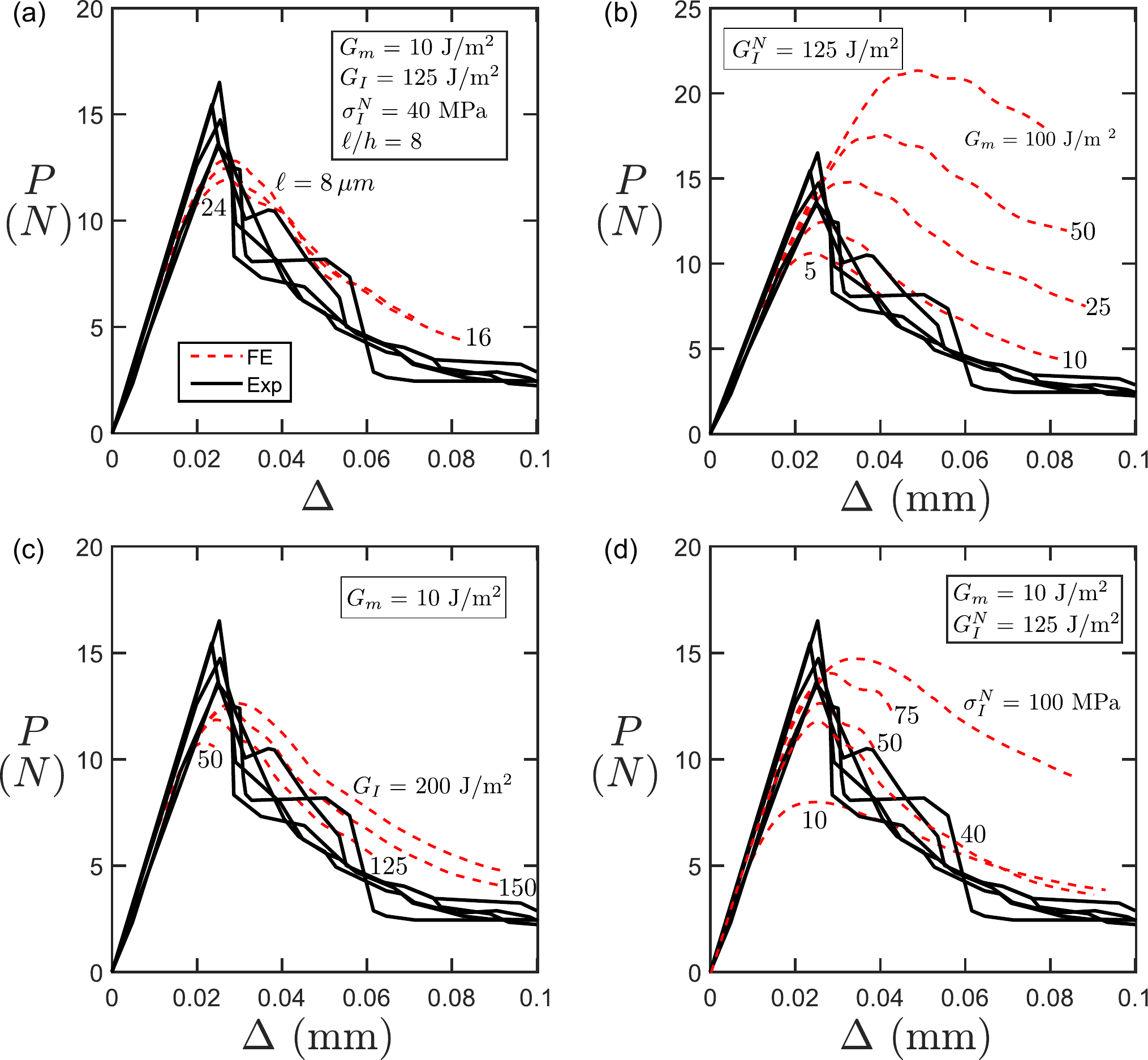}
    \caption{Sensitivity study of the parameters used in the simulations: (a) phase field length scale $\ell$, (b) fracture toughness of matrix $G_c$, (c) fracture toughness of interface $G_I^N$ and (d)interface strength $\sigma_N$. }
    \label{Fig.5}
\end{figure}

The fibre, matrix, and interface properties used in the previous section were taken as baseline values and a parametric study was carried out by simulating the mechanical response of the single notched beam bending test for different values of the phase field length scale $\ell$, matrix fracture energy release rate $G_m$, interface mode I fracture energy release rate $G_I^N$ and interface normal strength $\sigma_I^N$. The load-CMOD responses of these parametric analyses are plotted in Fig. \ref{Fig.5}. Fig. \ref{Fig.5}a shows that reducing the value of $\ell$ elevates the force-displacement response; in all cases, a constant ratio $\ell/h=8$ is adopted to ensure mesh independent results. This can be rationalised by recalling the relation between the phase field length scale and the material strength: $\sigma_c=\sqrt{27EG_c/(256 \ell)}$ (see, e.g., \cite{CMAME2018}). However, the influence of $\ell$ appears to be small. In agreement with fracture mechanics, phase field predicts a strength dominated behaviour (i.e., sensitive to the choice of $\ell$) when the initial defect is smaller than the transition flaw size, and a fracture dominated response (i.e., governed by $G_c$) for larger cracks \cite{Tanne2018}. In elastic-plastic materials, cracking always takes place at $G=G_c$ if the initial flaw is sufficiently large but the dissipation (R-curve) is influenced by $\ell$ \cite{JMPS2020}. As expected, both the peak load and the CMOD at the maximum load increase with the increasing $G_m$, see Fig. \ref{Fig.5}b. However, the interface fracture toughness has a relatively small effect on the load-CMOD responses, see Fig. \ref{Fig.5}c. Therefore, to enhance the overall fracture toughness, increasing the matrix toughness is more effective than increasing the interface fracture toughness. This supports the trend of using thermoplastic materials for high fracture toughness applications \cite{Tan2016}. The interface normal strength $\sigma_I^N$ has a significant impact on the maximum load, which correlates to the initiation of fibre-matrix interface debonding. From the above analysis, we can conclude that interface strength $\sigma_I^N$ determines the peak load, while both the matrix and the interface fracture toughness contribute to the softening behaviour of the overall mechanical response. 

\subsubsection{The effect of porosity} 

\begin{figure}[H]
    \centering
    \includegraphics[width=1.0\textwidth]{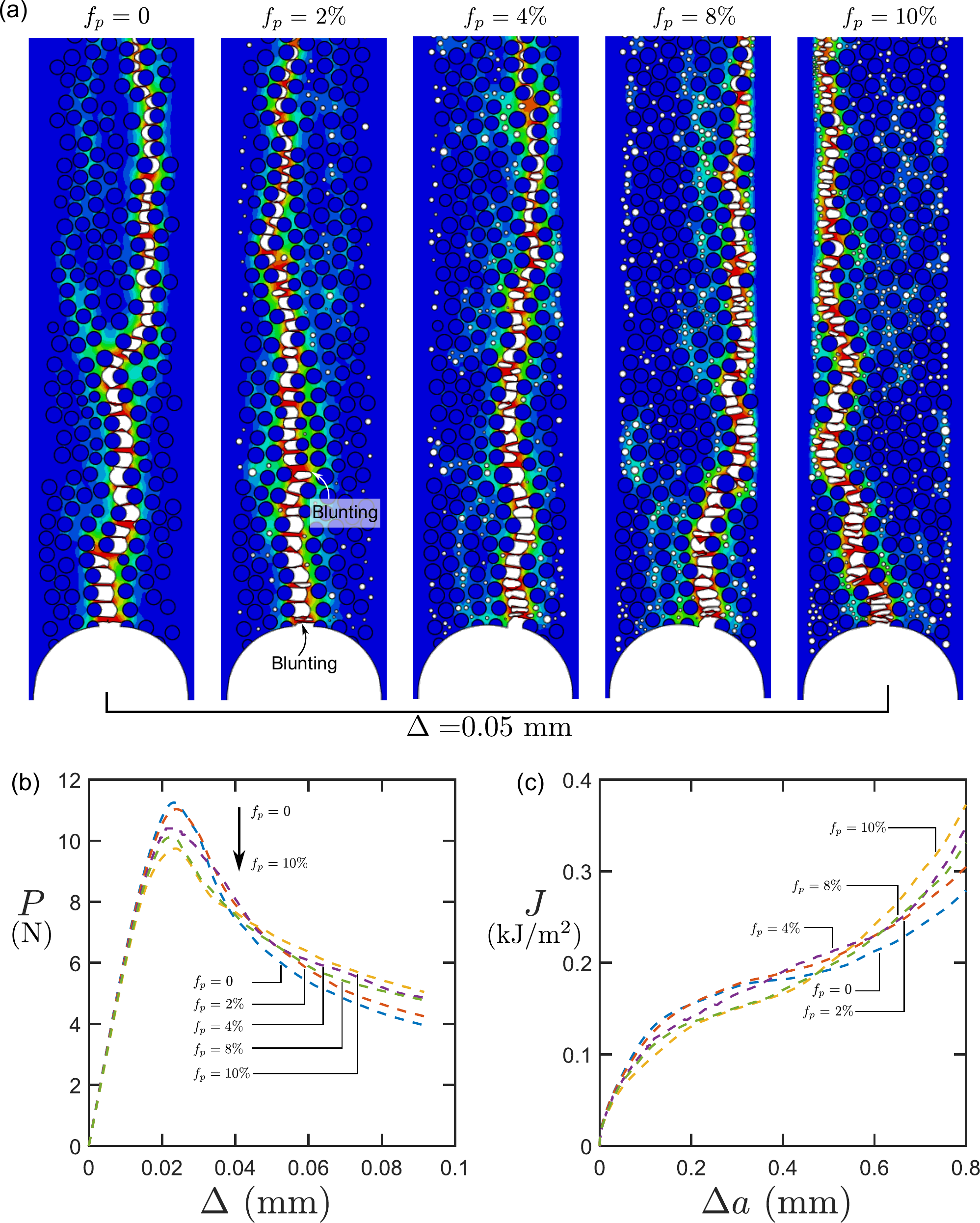}
    \caption{The role of porosity on: (a) the crack trajectory, (b) the load-CMOD response, and (c) the fracture resistance R-curves.}
    \label{Fig.6}
\end{figure}

After validating our model against the experimental results, we shall now proceed to explore the effect on the fracture behaviour of other microstructures, such as those arising from an increase in porosity or voids. In a composite material, a void is a pore that remains unfilled with polymer and fibres. Voids are typically the result of poor manufacturing of the material and are generally treated as defects as they can degrade matrix-dominated properties such as interlaminar shear strength, transverse tensile strength and longitudinal compressive strength, hence affecting the overall mechanical properties. The effect of porosity on strength has been assessed by Vajari \textit{et al.} \cite{Vajari2014}. Here, we quantify the influence of voids on both strength and fracture toughness. To achieve this, we introduce pores on the baseline model, with the porosity ranging from $f_p=$ 2\% to $f_p=$ 10\%. The porosity is represented by circular voids within the matrix and all the other conditions are kept the same. 2D models can provide quantitative insight into the role of porosity as pores in unidirectional ply have a tubular shape \cite{Hernandez2011}. The resulting crack trajectories for selected porosity levels are shown in Fig. \ref{Fig.6}a. Crack blunting was observed during the fracture process. The crack paths appear to be very sensitive to the porosity level. In addition, as shown in Fig. \ref{Fig.6}b, both modulus and strength decrease with increasing volume fraction of porosity. The strength is reduced by approximately 17\% in the presence of 10\% porosity. A similar degradation was measured by Olivier \textit{et al.}  \cite{Olivier1995} and predicted by Vajari \textit{et al.} \cite{Vajari2014}. Figure \ref{Fig.6}c shows how the R-curve of fibre-reinforced composites changed from `flat'-type to `rising'-type with increasing porosity. For the sample with higher porosity (10\%), the fracture toughness rises continuously with crack advance, exhibiting a more stable crack growth. The sample with 10\% porosity has a 37\% higher fracture toughness compared to the sample without porosity for $\Delta a=0.8$ mm. This toughening effect can be attributed to the circular holes that blunt the crack-tip and increase the fracture toughness \cite{Liu2020}. This finding differs from the effect associated with manufacturing induced defects, where voids degrade the mechanical behaviour \cite{Tan2018}. It should be noted that manufacturing-induced voids are commonly not regular and are more likely to be located close to fibre-matrix interface; hence reducing the interface and the macroscopic fracture toughnesses. For this virtual test case, all the voids have a regular circular shape and are located at the matrix pocket. Therefore, crack blunting effects are enabled. 

\section{Conclusions}
\label{Sec:ConcludingRemarks}

In this work, we present a novel coupled phased field and cohesive zone model to explore the effect of microstructure and constituent properties on the macroscopic fracture toughness. Several boundary value problems of particular interest are modelled to showcase its capabilities and gain physical insight.

First, an analysis of simple single-edge cracked plate tension tests on fibre-reinforced composites suggests that a weak fibre-matrix interface strength will raise the fracture toughness but reduce material strength. Secondly, the model is validated against single-edge notched beam bending experiments. Our predictions exhibit an excellent correlation with the experimental results both qualitatively and quantitatively. Subsequent parametric analyses suggest that increasing the matrix toughness is a more effective toughening mechanism than enhancing the interface fracture toughness. Finally, the influence of different microstructures with varying porosity levels is subsequently investigated to determine optimal toughening strategies. We show that introducing a volume fraction of void inclusions in the matrix-resin regions can enhance the composite fracture toughness due to crack blunting effects.

This embedded cell-based, combined phase field and cohesive zone computational framework provides a compelling multiscale virtual tool to investigate the role of the microstructure and material properties. This will lead to more efficient and rapid designs for enhancing the fracture toughness of energy-absorbing materials and structures.

\section{Acknowledgements}
\label{Sec:Acknowledgeoffunding}

W. Tan acknowledges financial support from the European Commission Graphene Flagship Core Project 3 (GrapheneCore3) under grant No. 881603. E. Mart\'{\i}nez-Pa\~neda acknowledges financial support from the EPSRC (grants EP/R010161/1 and EP/R017727/1) and from the Royal Commission for the 1851 Exhibition (RF496/2018).






\bibliographystyle{elsarticle-num}
\bibliography{library}


\end{document}